# Data traffic load balancing and QoS in IEEE 802.11 Network: Experiemental study of the SNR Effect

Adel SOUDANI [1,3], Thierry DIVOUX [2] & Rached TOURKI[1]

(Emails: adel.soudani@issatso.rnu.tn, Thierry.DIVOUX@cran.uhp-nancy.fr, rached.tourki@fsm.rnu.tn)

[1] Laboratoire d'Electronique et de Micro-Electronique (EµE), Faculté des Sciences de Monastir, 5019 Monastir, Tunisia

[2] Centre de Recherche en Automatique de Nancy (CRAN - UMR 7039), Nancy-University, CNRS, BP239, 54506 Vandoeuvre, France

[3] Middle East College of Information Technology. P.O. Box:79, P.C: 124. Al-Rusayl, Sultanate of Oman.

Corresponding author: Dr. Adel SOUDANI

**Abstract:**

Recently there is a growing interest in the internet and multimedia wireless networking where the bandwidth and the Quality of Service ( QoS) should be managed with careful. The performances of applications built on these systems depend on bandwidth availability that arises as the most limit of this solution. In fact, the IEEE 802.11 standards do not provide performed mechanism of loading distribution among different APs of the network. Then, an AP can be heavily overloaded leading to station throughput degradation.

For this problem, load balancing algorithms (LBA) was been considered as one of the attractive solution to share the traffic through the available access points (APs) bandwidths. However, applying the load balancing algorithm (LBA) and shifting a mobile connection from an access point to another without considering the signal to noise ratio (SNR) of the concerned APs might causes worst communication performances. This paper is then a contribution to check the performance's limits of the LBA algorithm through an experimental study of communication metrics for MPEG-4 video transmission over IEEE 802.11 network. The paper focuses, then, on the proposition of a new structure of the LBA algorithm with the introduction of the SNR level.

## 1. Introduction.

Nowadays, Hotspots environment based on IEEE 802.11 networks are widely used due to their low-cost hardware infrastructure. The growing interest in the internet and in multimedia wireless networking in these hotspot environments has revealed the necessity to focus more on the quality of service management [1,5]. In the work presented in [2], we have studied the different mechanisms of quality of service, being adopted in the IEEE 802.11 based networking systems. We have defended in this paper a general approach based on a centralized architecture intended to share mobiles on available APs that belong to the hotspot environment. The proposed approach is based on the use of the Load Balancing Algorithm (LBA) [2, 3, and 4]. The basic idea of the LBA algorithm is to ensure a fair (balanced) distribution of the traffic. For that purpose, if a new mobile access to the network or when a mobile change the BSS or when the application level change the requirements the LBA algorithm will be processed to find the best new distribution based only on traffic load criterion for fair bandwidth use. The load balancing algorithm (LBA) will evaluate the traffic load of each access point in the network to decide whether it is considered as under-loaded or overloaded. According to the LBA server decision, a new distribution of mobiles will be applied. Then, some mobile's connections will be shifted from AP to others, regardless the quality of their channel links, in the aim to share the general bandwidth of the network, We presented in [2] an efficient centralized architecture for the application of this approach. We designed the a protocol structure to manage data transaction between the LBA server, APs, and mobiles. A set of required new requests for quality of service handling in this architecture has been defined and checked with SDL tool. In addition, a simulation study was been carried using Opnet tool to testify the performances of our approach when applied to a network with different data flow types. The results obtained in this study proved the efficiency of this approach to enhance the quality of service at the receiver application level. It shows that, with the LBA application, the global dropped data amount at the network was minimised, the MAC access delay has been enhanced and the average network throughput to the applications' layers was increased.   . Despite the good performances proved  with the application of the  ,load balancing algorithm, our work proposed in [2] had not addressed the limitations of the LBA approach. In fact, when finding the optimal distribution, the LBA server doesn't take in account the SNR of the new  target AP  and no comparison with the SNR of the current AP will be performed.  So, it may happen that a mobile will be moved from an AP to another one that provides less SNR level that might destroy the quality of service. Taking in account the Shannon formula about the bandwidth effectively offered ($C_{max}$) by an access point that is expressed by equation 1.

$$C_{max} = BW * \log_2 (1 + SNR) \qquad (1)$$

Where BW is the bandwidth and SNR is the signal-to-noise ratio. Based on that we think that more work is yet required to explore the application's limits of the LBA approach. The impact of the SNR level of the wireless channel link to the efficiency of the LBA application should be clarified in order to correlate the involving of the SNR with the traffic load consideration in a new structure of the Load Balancing Algorithm (LBA).

This paper contributes in this objective at three levels.. It aims first of all to make a proof with an experimental study on the efficiency of the LBA algorithm to improve the quality of service not only in the network level, as being studied in [2], but also at the quality of service at application's processes. At a second level, this paper explores the application's limits of the LBA algorithm to enhance the quality of service in a communication process in regard to SNR of the wireless channel between the mobile station and the AP.. For this purpose, this paper presents a set of experimentation concerning the analysis of the LBA effect on the quality of service with the consideration of the associated SNR connection between the mobile and the access point AP. The main goal of this third part is to involve the SNR as a parameter in the LBA algorithm decision. In that case, the shifting of a mobile to a new connection with another AP will be applied not only with the traffic balancing consideration but also with the SNR value. Then, based on the obtained results the paper presents a new structure of the LBA algorithm.

2. **Overview of QoS in IEEE 802.11 and related work**

The basic criterion to establish a connection for a mobile in overlapped IEEE 802.11 cells is the level of the Signal to Noise Ratio (SNR). This criterion is based on the Shannon formula expressing the available bandwidth as an increasing function of the SNR value. As a consequence, a mobile station is associated to the access point offering the best Signal by Noise Ratio (SNR) independently of the load being applied to the access point by other users. This can cause, in many cases, unbalanced load between access points. Some access points will be over loaded, others are under loaded. When an access point is overloaded, applications requirements are not fulfilled and the end users will not be satisfied about the ensured quality of service (QoS) [2, 4]. To better manage quality of service in these networks, different mechanisms have been proposed to enhance the communication processes. We cite here some of the most important [5].

- **Service Differentiation.** Basically, service differentiation is achieved by two main methods: priority and fair scheduling [6]. While the former binds channel access to different traffic classes by prioritized contention parameters, the latter partitions the channel bandwidth fairly by regulating wait times of traffic classes in proportion according to given weights [5]. Used parameters for both approaches are Contention Window (CW) size, Backoff algorithm and Inter Frame Space (IFS). The main service differentiation mechanism is the 802.11e standard [7]. An access method called Hybrid Coordination Function (HCF) is introduced. It is a queue-based service differentiation that uses both DCF and PCF enhancements. HCF describes some enhanced QoS-specific functions, called contention-based HCF channel access and polling-based HCF access channel. These two functions are used during both contention and contention free periods to ensure QoS. Enhanced DCF (EDCF) is the contention-based HCF channel access. The goal of this scheme is to enhance DCF access mechanism of IEEE 802.11 and to provide a distributed access approach that can support service differentiation. The proposed scheme provides capability for up to eight types of traffic classes. It assigns a short contention window to high priority classes in order to ensure that in most cases, high priority classes will be able to transmit before the low-priority ones. For further differentiation, 802.11e proposes the use of different IFS set according to traffic classes. Instead of a DCF IFS (DIFS), an Arbitration IFS (AIFS) is used. Classes with smallest AIFS will have the highest priority.

- **Admission Control and Bandwidth Reservation.** Service differentiation is helpful in providing better QoS for multimedia data traffic under low to medium traffic load conditions. However, due to the inefficiency of the IEEE 802.11 MAC, service differentiation does not perform well under high traffic load conditions [5]. In this case admission control and bandwidth reservation become necessary in order to guarantee QoS of existing traffic. These two approaches are quite difficult to build up due to the nature of the wireless link and the access method. Admission control schemes can be roughly classified into measurement-based and calculation-based methods. In measurement-based schemes, admission control decisions are made based on the measurement of existing network status, such as throughput and delay. On the other hand, calculation-based schemes construct certain performance metrics or criteria for evaluating the status of the network [5].

- **Link Adaptation.** 802.11 specify multiple transmission rates but it intentionally leaves the rate adaptation and signalling mechanisms open. Since transmission rates differ with the channel conditions, an appropriate link adaptation mechanism is desirable to maximize the throughput under dynamically changing channel conditions [5]. Most link adaptation mechanisms focus on algorithms to switch among transmission rates specified in the Physical Layer Convergence Procedure.

All of these mechanisms deal with the quality of service at the connection level between the mobile and the next hop. However, the quality of service heavily in the global network state (number of connection by hop and traffic load), therefore different other research work had addresses the traffic load balancing approach in wireless LAN as a potential key to enhance its global performance.

- **Load Balancing approach and related contribution.** The work in [8] proposed a dynamic load balance algorithm to maximize the average Received Signal Strength and minimize the amount of associated stations per access point. In [4] Bianchi et al, proposed some packet level load balancing mechanisms based on the number of calls admitted in a cell. In [3], authors proposed a distributed load balancing architecture based on the access point's load. Over loaded access points are not authorized to associate new coming stations. The state of an access point (over loaded or not) is compared only to the neighbors AP. The defined algorithm can lead to a non optimized number of dissociations and associations. The authors of [1] describe a centralized load balancing algorithm which is not completely transparent to wireless users: stations have to explicitly change positions to get the required quality of service. A scheme proposed in [9] aims to fairly distribute load between access points by changing their coverage area. Some other works describe load balancing mechanisms for 802.11 wireless networks but are applied in particular contexts such as in [10] which study a routing connection admission control in ESS (Extended Service Set) mesh networks.

In [11], Al-Rizzo et al proposed an algorithm to decrease congestion and to balance the user's traffics in IEEE 802.11. Each access point has one different AP channel to avoid congestion event and signal interference problems. The algorithm found the Most Congested Access Point (MCAP), analyses the users association and decrease the congestion of MCAP. A user will not be connected to an AP with the result that the Signal to Interference Ratio (SIR) is positive and the signal power is more than a fixed threshold.

[12, 15] Presented a Cell breathing Technique to balance the APs load and improve QoS of real time applications. It reduces the signal length of the congested APs (cell area), so reduce the AP's impact and user's number per congested AP. On the other hand, it increase the signal length and the impact of the under loaded APs. Then, as a result, it makes new connection with the disconnected stations.

T. Scully et al in [14] had studied the efficiency of Genetic-based load balancing algorithm. They proposed, for this purpose, two algorithms: the first one is a standard genetic algorithm (MacroGA), while the second is a micro-genetic algorithm ( MicroGA). They evaluated these algorithms using an empirical approach and they demonstrated that both algorithms outperform the WLAN techniques in terms of total network throughput and distribution of user bandwidth. However, the studied approach doesn't consider the signal strength in the mobile distribution among the available access points.

3. **The proposed architecture supporting the LBA for QoS enhancement**

The architecture we have proposed to support the application of the LBA algorithm is a centralized architecture around a Load balancing server able to provide the best mobile distribution any time a new event will occur in the network. We have proposed for this architecture a protocol structure for the set of exchanged request making up the different steps of signaling, connection association and dissociation between mobiles and APs [2].

Even though that, in general, the centralized architecture may not be the suitable one for communication environment with user QoS constraints since it leads to a system with one point failure, this architecture is the most adequate to the LBA application. In fact, to find the best new fair mobile station distribution the

LBA algorithm requires a collection of the different AP traffic load states. In a distributed approach for LBA application, the decision will take more processing time and requires more interactions between AP that increases the traffic load for APs.

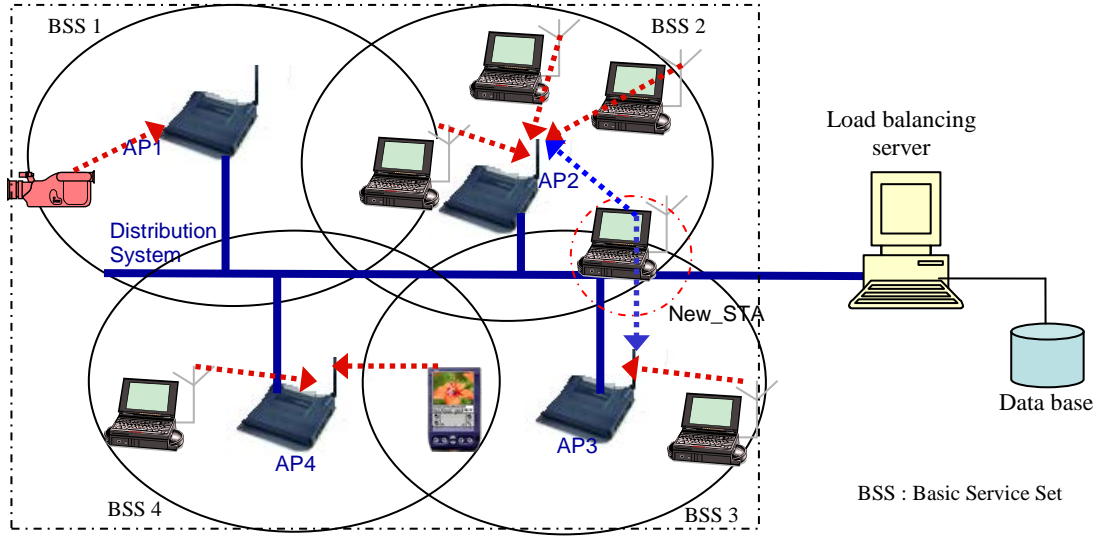

**Figure 1: The proposed IEEE 802.11 target architecture supporting the LBA algorithm**

The LBA can be summarised as follow, the algorithm checks if the new distribution is balanced mainly by computing a balance index β. The balance index appeared in the first time in [13] and it is used in [1, 3] as a performance measure. The balance index reflects the used capacity in each access point. It is computed in each overlapping zone between two or more APs in the WLAN Extended Service Set.

Let Ti be the total traffic of the APi. Then, the balance index $\beta_j$ of an overlapping zone $Z_j$ is:

$$\beta_j = (\Sigma_i T_i)^2 / (n * \Sigma_i T_i^2) \qquad (2)$$

With Ti is the total traffic of an APi overlapping with other access points in the zone j and n is the number of access points overlapping in the zone j.

The proposed distribution of mobile stations is balanced if the balance indexes of all the overlapping cells converge to 1. At this step, overlapping Access Points in the Wireless LAN are fairly loaded. We use a tolerance parameter α that defines the balancing zone width. We use therefore the Average Network Load (ANL) calculated as:

$$ANL = \Sigma_i T_i / n \qquad (3)$$

An AP is considered overloaded if its load exceeds a certain threshold as expressed in equation 4

$$\delta_1 = ANL + \alpha.ANL. \quad (4)$$

$$\delta_2 = ANL - \alpha.ANL \quad (5)$$

It is balanced when its load is in the interval [$\delta_1$, $\delta_2$], An AP is under loaded if its load is under $\delta_2$ (equation 5). The $\alpha$ parameter acts upon the stability of the system, when its value go near zero an access point state will topple during the LBA application between over-loaded and under-loaded state. In this case it will not be easy to find the best distribution and the algorithm may not converge. The compromise that should be considered during the definition of the $\alpha$ value is to get a sufficient gap within the AP will be considered loaded, but not too much to get unfair traffic load distribution. In [2] we have used 20% as value of $\alpha$, but we think that it can range between 10% et 20% according to the network and the application constraints.

When a new mobile station asks for a connection with an access point, this later will check if it will be able to provide the required bandwidth for the user application. If possible the connection request will be accepted and the LBA server will not be invoked, avoiding network instability. In the other case the load balancing algorithm calculates the average network load, the $\delta_1$ and $\delta_2$ balancing thresholds for all the APs in the network. If there is some overloaded access point, the algorithm has to seek for a new distribution of mobile stations to have balanced load among access points. Mobile stations have to be moved from overloaded access points to less loaded ones overlapping in the zone$_{min}$. Zone$_{min}$ is the overlapping zone which has the minimum balance index value. The load balancing algorithm computes the difference between the load of the most loaded AP and the average network load. The station to move is the one with the load nearest to this difference.

As being explained in the previous section, in the LBA algorithm, the dissociation between a mobile and an access point is taken with the only consideration of APs load balancing. The moving decision doesn't care about the comparison between source and destination AP provided SNR. So, even it had been proved that the LBA enhances the provided quality of service at the reception level it is clear that in its actual structure, the LBA suffers of some limits.

In this context, this paper presents an experimental study in order to check the efficiency of the LBA application and to discover its limits

4. **Experimental target platform and QoS metrics**

The basic idea we had adopted to check the influences of the LBA algorithm application is summarised in figure 2.

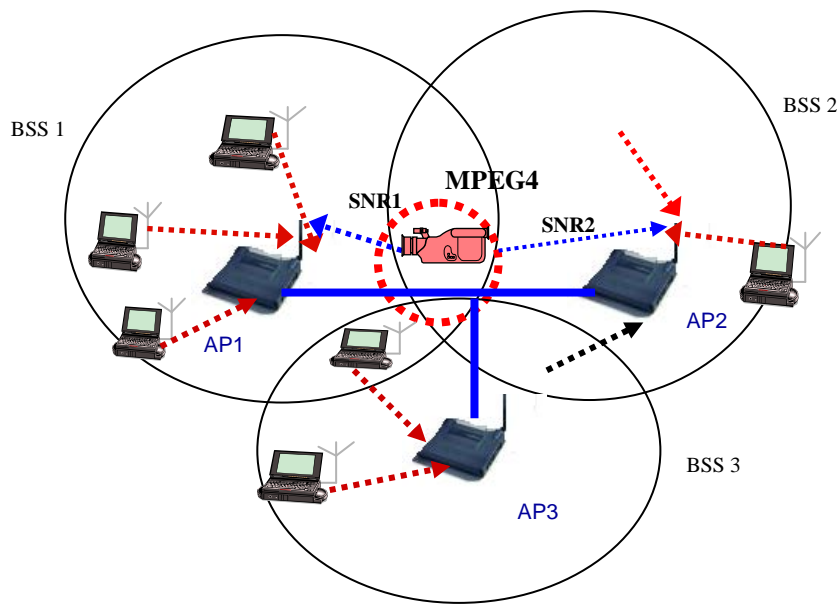

**Figure 2.** Experimentation of the LBA algorithm application

In this architecture some communication processes have been introduced to load the APs. These processes carry FTP and HTTP traffics. A specific MPEG4 short head video process has been, also, used to analyse the quality of service provided with a connection through AP1 and AP2. The connection with AP1 and AP2 are provided with different SNR values. The traffic load of an AP is being calculated as the average sum of its up-link and the down-link traffic.

To make possible the analysis of the quality of service ensured for a communication process, we had developed a specific Java-client application. This application (QoS monitoring) is able to measure a set of parameters that can be used to quantify the quality of service provided for the communication process at the receiver level.

The application receives the data flow from the appropriate communication channel of the wireless video camera (*Trend net IP 300 w*). This camera incorporating an embedded video server, provides MPEG-4 short header coded data structure with different resolutions and different image's rates (1, 3, 7, 15, 25 images/s). As, a standard way to use this camera, the provider makes easy to capture the video through internet navigator using HTTP protocol. However this method doesn't allow the user to analyse the different parameters of video quality, since he cannot add his appropriate function to make possible analysis of the provided video streaming.

Thus why, we have developed this application that pops the compressed video *( VAM format)* form the memory of the server. Several metrics have been defined in order to quantify the quality of the received video stream. The analysis of the received flow is processed in real time (figure 3). So, the developed application calculates these values and plots the variation of the defined metrics during data reception.

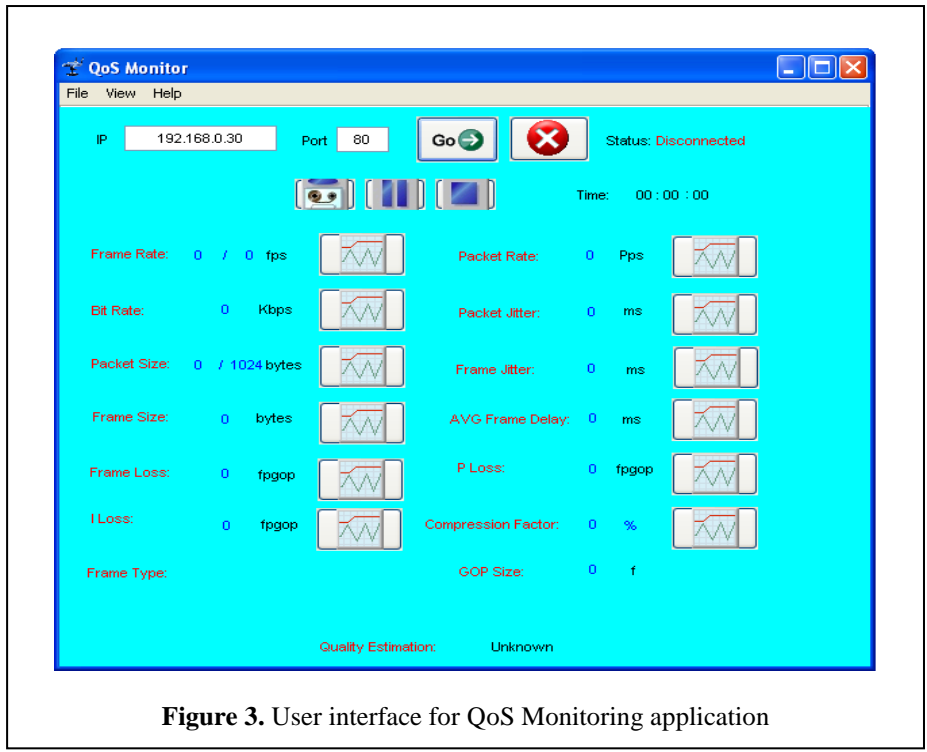

**Figure 3.** User interface for QoS Monitoring application

### 4.1 General features of the processed video stream

As being previously explained, to evaluate the efficiency of the LBA application in regard to the provided SNR value and the traffic load of the associated AP, we have used the *QoS Monitor* application that allows to measure in a real time way different metrics.

### 4.2 Used QoS metrics,

As being explained in the previous section the QoS monitoring application provides a set of metrics. In this paper, we will present only the results for some metrics that we consider as the most significant to quantify the quality of service at the reception level: All of these metrics will be applied to the MPEG 4 short head process at the reception level in order to quantify the quality of service for the end user.

- **The bit rate**, it represents the most used criterion to measure the quality of service of a given communication process.

- **The end to end delay,** it reflects the capacity of the system to satisfy constraints of short time transmission for time sensitive applications.
- **The Jitter,** the fluctuation of the inter-arrival time image at the reception level influences mainly the synchronisation. More is the fluctuation of the jitter worst is the quality of the received video.
- **Video PSNR** (*Peak Signal Noise Ratio*)**,** it represents the quality of the received video. It is calculated, in decibels, over all the video between source and received images. Its expression is given by equation 6

$$PSNR = 20 \log_{10} (255/RMSE), \qquad (6)$$

With RMSE (Root Mean Square Error)

The mean-squared error is represented by the following equation:

$$MSE = \frac{\sum_{M,N} [I_1(m,n) - I_2(m,n)]^2}{M * N} \qquad (7)$$

In the previous MSE equation, $I_1$ and $I_2$ represent pixel's values respectively for the source image and the received image. M and N represents the number of rows and columns in the images. The video PSNR is calculated is the average PSNR of N images received in a time period.

## 5. Experimentation and quality of service analysis of the LBA algorithm application

As a first goal of this study we have tried to make an experimental proof that the application of LBA algorithm enhances the quality of service. We have measured the defined metrics at the reception level for different traffic loads of the AP and for different SNR values.

Figure 4 plots the evolution of the bit rate of the MPEG4 process at the reception level in regard to the traffic load and the SNR value. We notice that for an access point with different traffic loads, the bit rate of the processed MPEG flow increases with the SNR. However, we can also see that the evolution of the bit rate depends on the current traffic crossing the access point. In fact, with SNR value of 50 dB for traffic load 12237 Kbps the bit rate (243kbps) provided by the AP for the MPEG4 process is less than the bit rate (457 Kbps) provided with 2944 Kbps as à traffic load. In addition, figure 4 shows that with SNR level at 20 dB we can reach a bit rate of 360 Kbps through AP loaded of about 570 Kbps, this bit rate level is much more better than the bit rate provided with 50 dB under a bit rate level of 12237 Kbps (234 kbps).

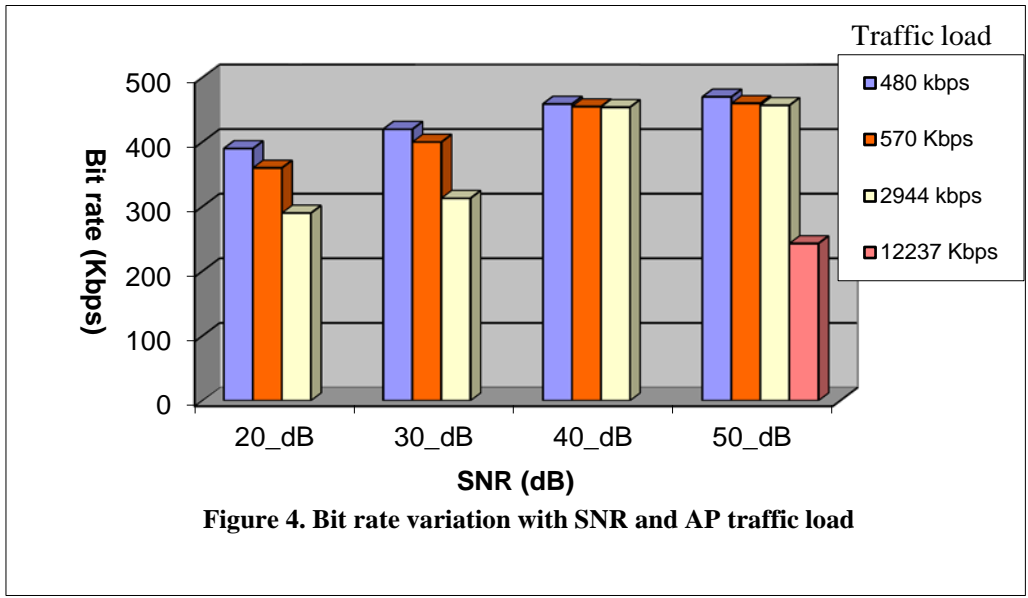

Figure 4. Bit rate variation with SNR and AP traffic load

We have measured, for the same experience, the end to end delay for image transmission between the wireless camera and the application running in a PC, through an AP. Figure 5 resumes the end to end delay according to the variation of the SNR and the AP load traffic. According to this graph, for all traffic loads, the image's end to end delay is conversely proportional to the SNR value. This statement confirms that the connection to AP with a high SNR value reduces the end to end delay. However, we note that for an AP traffic load of 12237 kbps with a high SNR value ( 50 dB), the image end to end delay (333 ms) is worse than the one measured ( 52 ms) for a traffic load of 480 kbps . This expresses that in general, the end to end delay, may not be improved through a new connection with high SNR value and heavy traffic load.

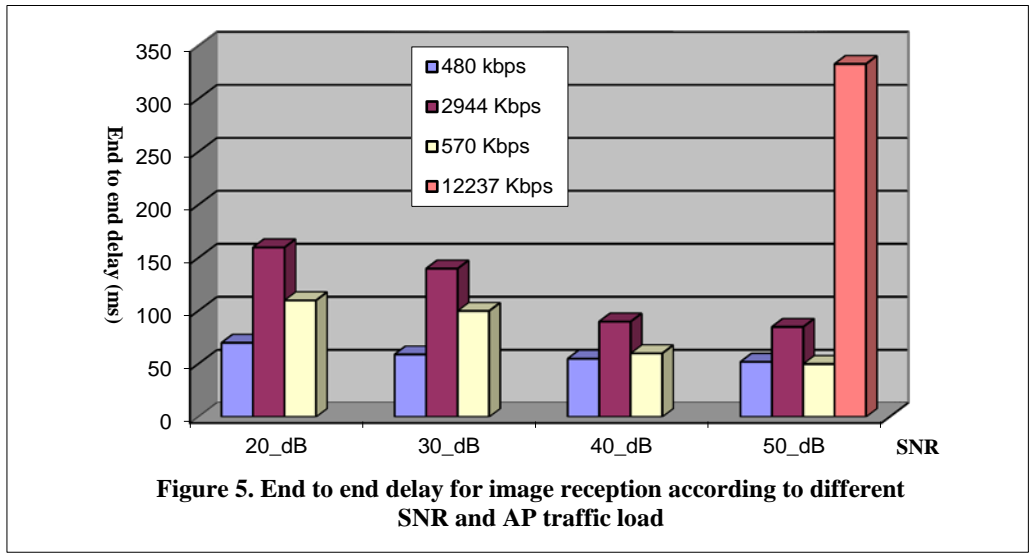

**Figure 5. End to end delay for image reception according to different SNR and AP traffic load**

Figure 6 sums up the average of the absolute value of the image jitter at the reception level for different SNR levels and traffic loads. We see, through this figure, that the average of the image jitter absolute value decreases for the same traffic load (570 kbps) when the SNR value increases. However, this figure shows that the interval of jitter variation for a traffic load of 12237 kbps is very high with a maximum value of 300 ms. This measured interval for jitter range is the widest of all the measured interval within the set of traffic load values{2944 Kbps, 570 Kbps and 480 Kbps} values. This result confirms the expected conclusion that, obviously, when we load the AP we get decrease in the quality of service. The increase on the Jitter range is a symptom of big depth of packet's buffering that increases the packet losses probability and the end to end delays. One other important note from this figure is that weak Jitter intervals was measured at 40 dB for all the values of traffic load when compared to the Jitter interval width measured for 12273 Kbps at 50 dB. This means that if we take only the SNR as a criterion, better quality of service can be ensured through connection with weak SNR level.

These results confirm the same observation. In fact, the jitter variation is not optimised only when we increase the SNR of Mobile/ AP connection. We should, in fact, consider the traffic load exchanged through this AP.

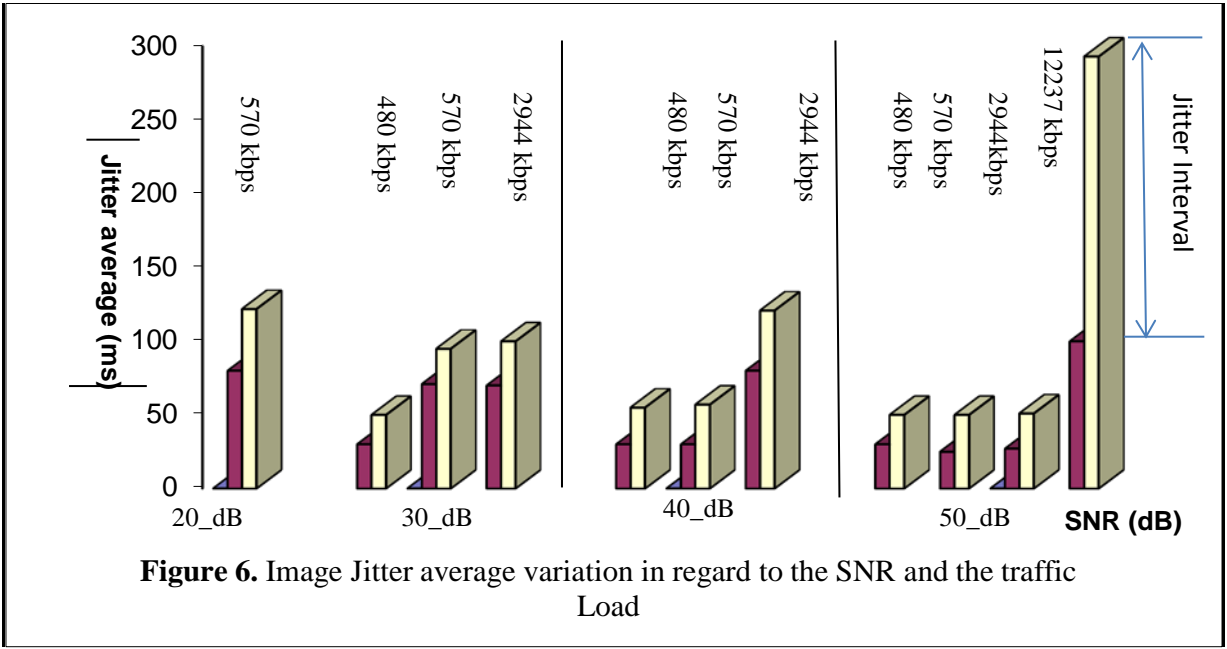

**Figure 6.** Image Jitter average variation in regard to the SNR and the traffic Load

The same note is confirmed for the quality of video received at the application level. Figure 7 summarises the MPEG4 quality with the PSNR values for different connection mobile-AP SNR levels and traffic loads.

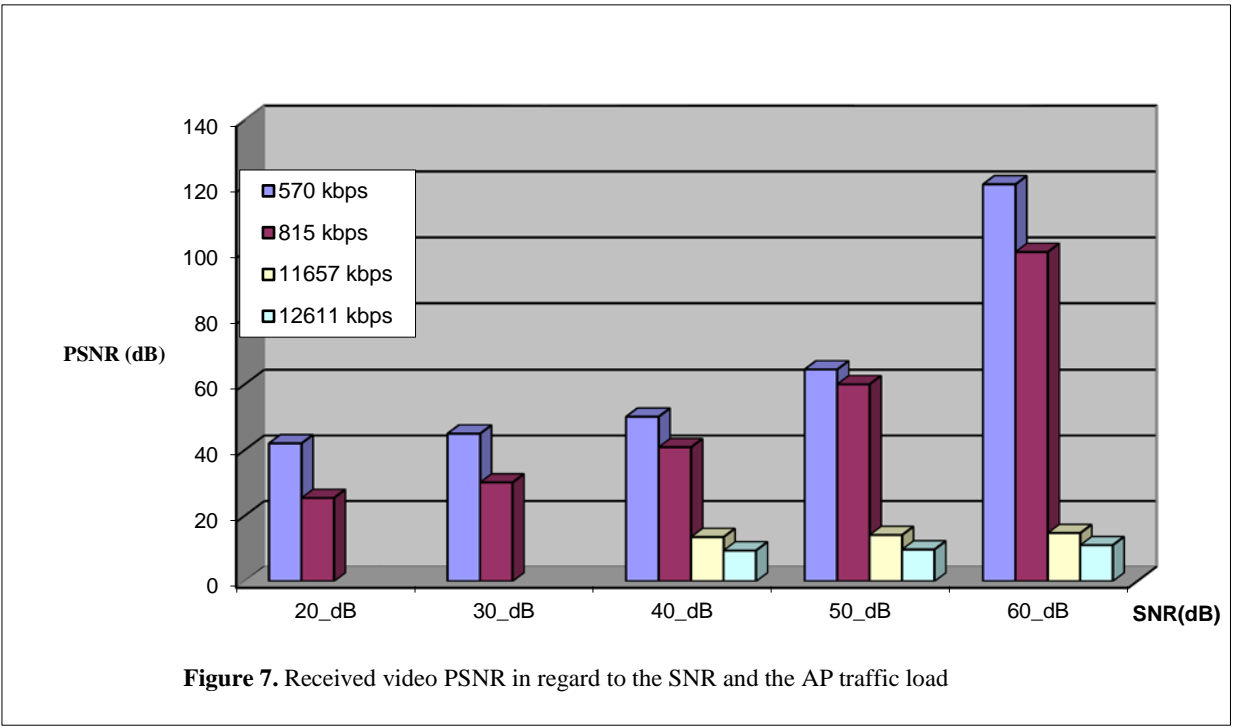

**Figure 7.** Received video PSNR in regard to the SNR and the AP traffic load

This figure (figure 7) shows that the PSNR quality of the received video will be enhanced, for the same AP traffic load (815 kbps), when SNR level for the mobile/AP increases. This is an expected result since we know that the wireless link bandwidth will be increased with the increase in SNR level. With higher bandwidth the transmission time for packet decreases which in turn decreases the collusion rate and then the packet losses probability. However, we see that for 30 dB SNR level, we got a PSNR received video (44, 9 dB) when the traffic load is 570 kbps. With 11657 kbps AP traffic load, we have measured for higher SNR level ( 50 dB) a PSNR at 14.1 dB. This note reveals that the PSNR will not be increased automatically in all cases when we increase the SNR level. In fact the load of the target AP has a big influence.

As a conclusion of this first experimentation set of results, we can conclude that the SNR level that characterizes the physical link between mobiles and AP is not enough to predict the quality of service that will be ensured over the AP. In fact, over all these experimentations, it had been proved that the AP traffic load influences heavily the provided quality of service for a given communication process. Consequently, during the application of the LBA algorithm, when moving mobile from an AP connection to another, aiming to distribute the traffic and to optimize the provided quality of service, we have to care about the SNR level provided by the destination AP.

**6.        Duality between load traffic and SNR level for quality of service insurance**

The question, at this level, is what should be the new proportions of the SNR level and the associated AP traffic load when taking the moving decision in the Load Balancing Algorithm (LBA) algorithm? For this purpose we have tried to evaluate through others experimentations the relation that should be considered for le LBA algorithm when sharing the mobiles through the available APs. We have used the same analysis platform to address this problem. So, while keeping the same communication process issued by the MPEG4 camera we have measured the quality of service for connection through two APs with different traffic loads. The same QoS metrics have been used to quantify the quality of the received video.  The idea of these new measurements is to evaluate the QoS metrics, for the MPEG4 short head process, when the wireless video camera connection is moved from a heavily loaded AP (AP_1) to a new AP (AP_2) with under-load traffic state. The new connection with the AP_2 is being evaluated with different SNR level. The main goal is to bring out an association between the SNR and the traffic load when making the decision about the dissociation of a mobile from an access point (AP_1) and its association with a new AP (AP_2).

Figure 8 resumes the bit rate evolution when the connection is first of all established with a heavily loaded AP ( AP_1) and then moved to less loaded AP (AP_2). Figure 8 (a) shows that for the same SNR level (80 dB), when the connection is moved to AP_2 ( 4050 Kbps) which is less loaded compared to AP_1 (2100 kbps) the bit rate increases. We see, also, that even when AP_2 has a SNR connection, with the wireless video camera, of 50 dB we get an increase in bit rate, which means that this note continue to be right for higher SNR level ({60…80} dB). The same reading is applied for figure 8 (b & c) using different starting SNR with AP_1 and also different traffic load for AP_1 and AP_2.

The general conclusion is that when the connection is moved from AP_1 to AP_2, that is under-loaded, it will increases the bit rate for this connection unless the SNR for AP_2 (SNR_AP2) is equal or under than (SNR_AP_1/2).

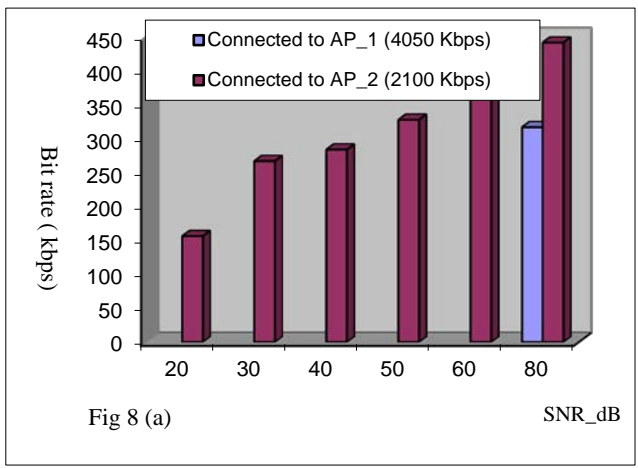

Fig 8 (a)

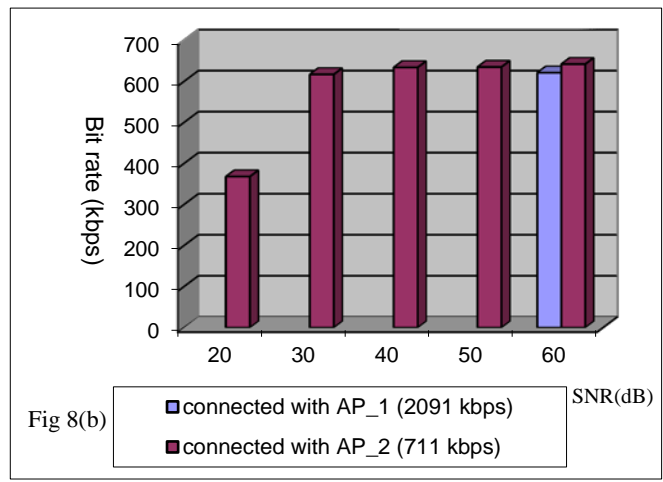

Fig 8(b)

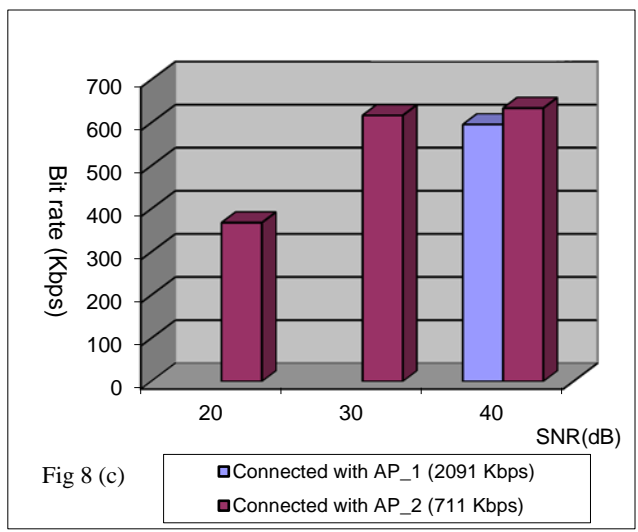

Fig 8 (c)

**Figure 8,** Bit rate evolution for MPEG 4 process when moving connection from heavily loaded AP to under-loaded one.

For the others metrics studied in this section, only one figure for each metric will be presented to resume its evolution in regard to AP traffic load and the SNR connection level.

Figure 9 represents the end to end image delay (ms) evolution when the connection is moved from AP_1 to AP_2. According to this figure we can see that the new connection with 80_dB with AP_2 will make short the end to end delay (0.096 ms instead of 0.13 ms). This enhancement in bit rate will not occur when the AP2 provides a SNR connection (SNR_AP2) for the mobile lower or equal to (SNR_AP1/ 2).

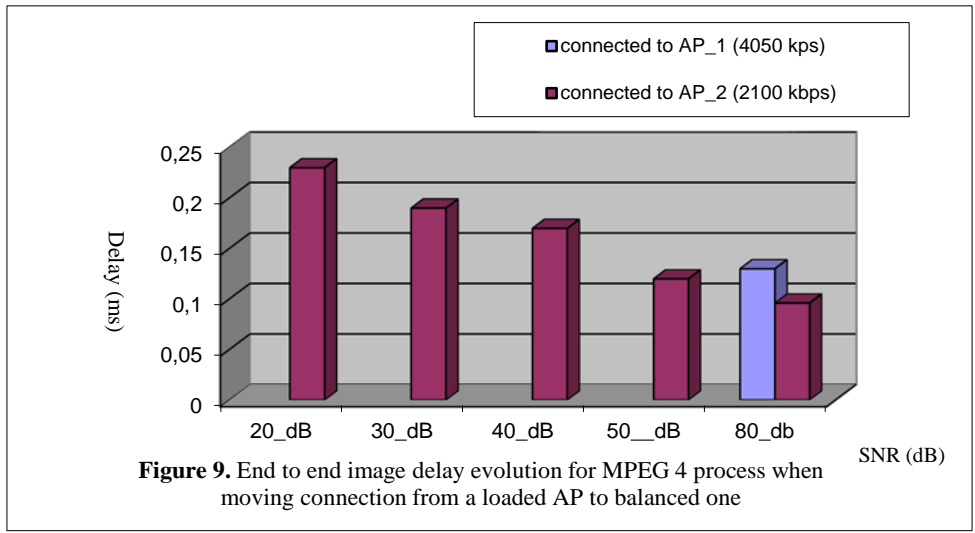

**Figure 9.** End to end image delay evolution for MPEG 4 process when moving connection from a loaded AP to balanced one

The quality of the received video is evaluated with the same scenario (figure 10). We note at this level the difficulty to get measurements for different SNR level and heavy traffic loads. This figure proves the same statement. In fact, the quality of the received video, evaluated with PSNR, will not be improved if the SNR connection with AP2 is not higher than (SNR_AP1/2).

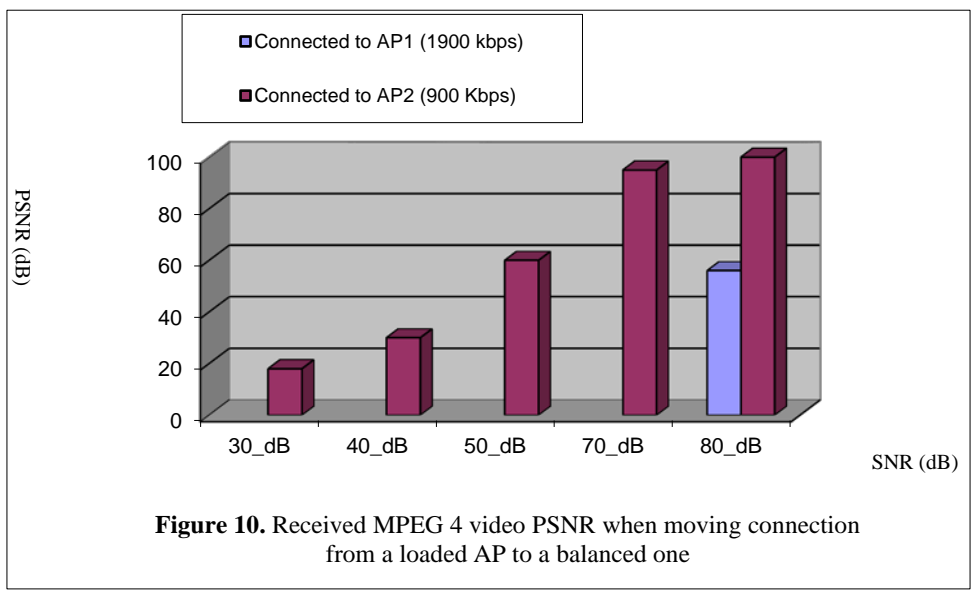

**Figure 10.** Received MPEG 4 video PSNR when moving connection from a loaded AP to a balanced one

7. **Enhanced new structure of the LBA algorithm**

Based in the analysis of the previous measurements for QoS metrics, we have concluded that the structure of the LBA algorithm should be modified to take in account, both the provided SNR level and the data traffic, during the dissociation and the association between a mobile and two APs. the objective of this study is to introduce a new criterion when we apply the LBA for a fair sharing of the mobiles through the available APs to enhance the achieved QoS. We remind that the LBA algorithm is processed in different steps

- **The first step,** it consists of identifying which AP has to dissociate mobiles and which one can receive new connection. At this level the algorithm is based on the ANL (Average Network Load) to determine for an overlapped area between two or more APs, those considered over-loaded and under-loaded. So, as being explained in section 3, an AP is considered balanced when its load is in the interval $[\delta_1, \delta_2]$ (figure 11).

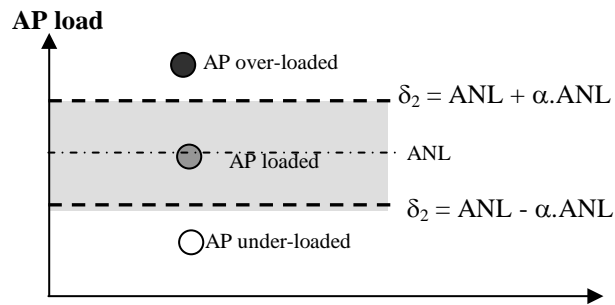

**Figure 11**. General consideration in AP traffic load classification

- **The second step** is to select from the overlapped area between two APs, the best candidate (mobile) to achieve the network balanced situation. In this step, we have to find the mobile that when moving it from the over-loaded AP to the under-loaded AP will contribute in the network traffic load balancing.

The main contribution we introduce in this algorithm is located at this step, in fact, we have proved, through the presented measurements of some QoS metrics, that these later will not be enhanced when moving the connection from AP to another unless the destination AP provides a SNR level at least higher or equal to the half of the SNR provided by the first AP. The new proposed decision in the LBA algorithm can be then summarized by figure 12.

- **Third step,** is related to the application of the final distribution by the load traffic distribution server to the set of access point in the network through the defined protocol architecture in [2].

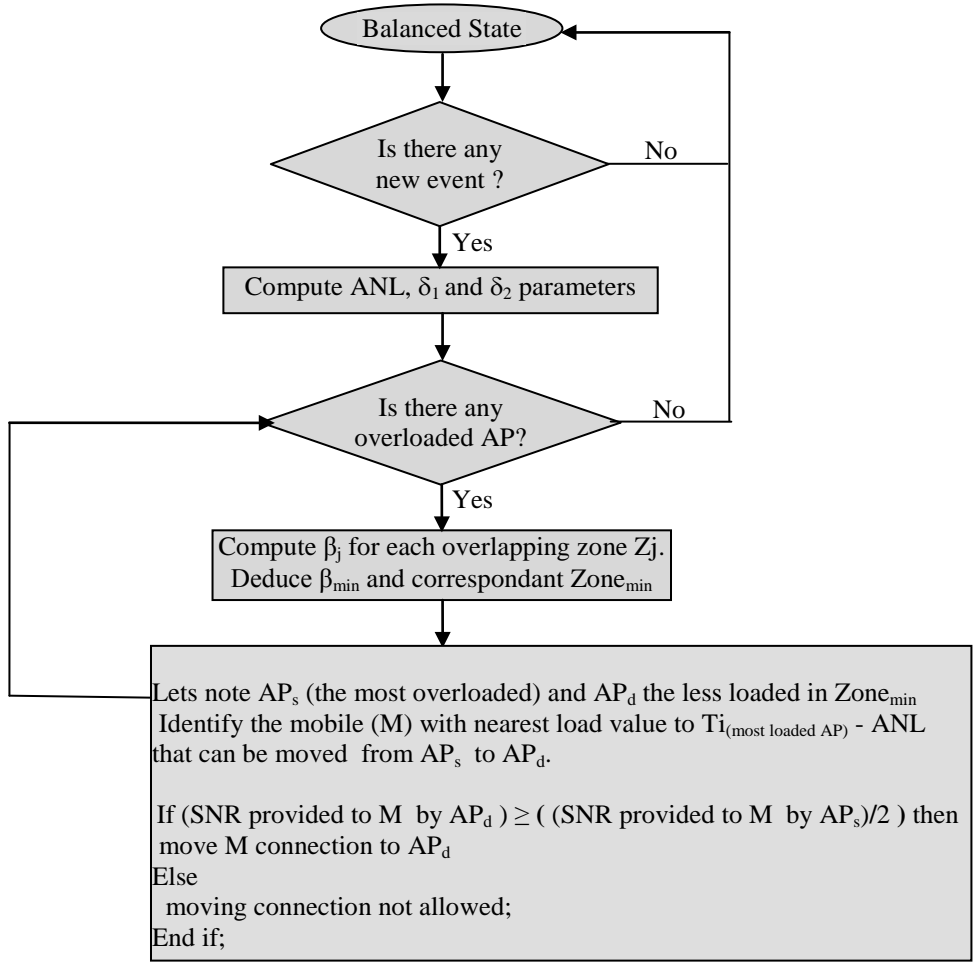

**Figure 12**. General consideration in AP traffic load classification

$Zone_{min}$ is the overlapping zone which has the minimum balance index value [1]

To evaluate the performances of this enhanced structure of the LBA algorithm we have applied it to a WLAN where we processed the wireless MPEG-4 video transmission within different others data communication processes based mainly on HTTP and FTP. The connection of the wireless camera (sender of MPEG-4 video process) has been shifted from AP- source (AP-S), considered heavily loaded and providing high SNR level, to AP-destination (AP-D) with less traffic load. We have considered during this study different offered SNR between the connection AP-D and the wireless camera in order to check up the capabilities of this new LBA structure to enhance the quality of service at the application level. Table 1 sums up the significant part of these results. Several QoS metrics have been evaluated for the connection with AP-S and AP-D. In this table some cells are labeled with 'NM' to express that it was impossible to get these values either due to the weak signal strength or for overloaded traffic through the AP.

Table. Performances evaluation of the LBA algorithm when considering the signal strength and the AP data traffic load

|  | Connected to AP1 overloaded before applying LBA SNR= 80dB | Mobile station connected to AP2, acceptor for new connections, after applying the LBA AP2 connection with different SNR levels and traffic load | | | | | | |
|---|---|---|---|---|---|---|---|---|
|  | Traffic load for AP1 (Kbps) | SNR (dB) | Traffic load for AP2 (Kbps) | | | | | |
|  | 4150 kbps |  | 450 | 1000 | 1200 | 2010 | 3000 | 11000 |
| Bit rate (kbps) | 495 | 30 dB | 490 | 480 | 300 | 286 | 200 | NM |
|  |  | 40 dB | 634 | 620 | 611 | 602 | 421 | 60 |
|  |  | 50 dB | 760 | 751 | 724 | 610 | 520 | 130 |
| Jitter (ms) | 36 | 30 dB | 028 | 035 | 036 | 036 | 037 | NM |
|  |  | 40 dB | 021 | 029 | 035 | 034 | 037 | NM |
|  |  | 50 dB | 021 | 029 | 035 | 034 | 036 | NM |
| Delay packet (ms) | 220 | 30 dB | 70 | 110 | 140 | 230 | 255 | 311 |
|  |  | 40 dB | 59 | 100 | 123 | 190 | 208 | 295 |
|  |  | 50 dB | 55 | 65 | 118 | 170 | 192 | 240 |
| PSNR (dB) | 20 | 30 dB | 43 | 41.2 | 19.2 | 18.2 | 17.5 | NM |
|  |  | 40 dB | 46.5 | 45.2 | 26.9 | 20.9 | 20.7 | NM |
|  |  | 50 dB | 52.1 | 51.8 | 33.4 | 22 | 21.9 | 12.2 |

We see, that for all the considered metrics, the assumption, that the QoS will be enhanced when the LBA establishes a new connection with an AP that offers SNR level higher or equal than the half of the pervious connection SNR level, is verified. It is shown for all the metrics when the SNR level with the AP-D is equal to 40 dB or 50 dB. We see that the obtained values reflect better quality of service than that evaluated at 80 dB with heavy traffic load. We also see, for some metrics, that when the traffic load of the AP-D is too much low, the new connection can achieve better quality of service even with weak SNR level (less than 40 dB). In fact with a new connection with AP-D 30 dB having a data traffic load of 450 Kbps the obtained end to end delay (70 ms) is better than the measured value of this delay at 80 dB. This note confirms that our proposal to enhance the LBA structure is valid when we are at the same range of data traffic load, which is often the case in the LBA application. In this table, the measures of the absolute jitter values don't provide a good conclusion. This is due to the fluctuated nature of the jitter.

As a conclusion, this experiment confirms that the new structure of this algorithm for load balancing in overloaded wireless network. It offers more reliability to enhance the quality of service in the network.

Several works had addressed the problem of load balancing and quality of service QoS in wireless environment. In these works, often, the traffic load and the signal strength had been considered independently. However, and as being shown in this paper, the best quality of service will be ensured when both of them are considered. Table 2 sums up a comparison of our solution in regard to other proposed solutions.

Table 2. Comparison with other's solutions

| | | Load balancing through channel [19] | Admission control with load balancing [17] | MITOS [18] | Inter-Access Point Communications [20] | Our solution |
|---|---|---|---|---|---|---|
| Wireless Network | | Wi-fi based networks | | | | |
| Proposed solution architecture | Centralized | | * | * | | * |
| | Distributed | * | * | | * | |
| Considered parameters for mobile station distribution | SNR/radio Strength | | | | * | * |
| | Time Space | | * | | | |
| | Distance | | | * | | |
| | Traffic load | | | | | * |
| | Channel | * | | | | |
| Approach for Performances evaluation | Simulation | * | * | * | * | |
| | Experimentation | | | | | * |

8. **Conclusion and future works**

In this paper, we have studied the enhancement achieved by the application of the LBA algorithm on the quality of service at the application level in IEEE 802.11 network. For this purpose, we have based our study on the analysis of some QoS metrics (Jitter, end to end delay, bit rate and PSNR) for a video MPEG-4 process transmission via different APs in the same network.

As a first result, we have proved that the application of the LBA algorithm to share fairly the traffic through the available APs in the network will contribute in the enhancement of the QoS of the traffic at the reception level.

As a second goal of this paper, we have addressed the relationship that should be considered between the provided SNR and the traffic load by the target AP and its comparison to those that characterize the current AP. We have conclude that a relationship should be filled between SNR levels of the source AP and the destination AP when moving a mobile connection as a decision of the LBA algorithm. Based on these results we have proposed a new structure for the LBA algorithm.

For the future work, we would like to extend the LBA structure in order to introduce others techniques that have been studied in related works such as cell breathing and channel allocation. In addition, further study should be carried to analyze the performances of centralize solution and distributed solution for the application of the LBA algorithm.